\documentclass[dvips,12pt,a4paper]{article}
\usepackage{epsfig}
\usepackage{a4}
\usepackage{amsmath}
\usepackage{latexsym}
\begin{document}
%%%%%%%%%%%%%%%%%%%%%%%%%%%%%%%%%%%%%%%%%%%%%%%%%%%%%%%%%%%%%%%%%%%%%%%%
\def\vecE{{\pmb{E}}}
\def\vecs{{\pmb{\sigma}}}
\def\vecH{{\pmb{H}}}
\def\d{{\rm d}}
\def\reg{{\rm reg}}
 \thispagestyle{empty}
 \begin{flushright}
 {\tt University of Bergen, Department of Physics}    \\[2mm]
 {\tt Scientific/Technical Report No.1996-04}    \\[2mm]
 {\tt ISSN 0803-2696} \\[5mm]
 {hep-ph/9606298} \\[5mm]
 {June 1996}           \\
\end{flushright}
\vspace*{3cm}

\pagenumbering{arabic} 

\begin{center}
{\bf \large Dyon mass bounds from electric dipole moments}
\vspace{1cm}

{\bf Per Osland}\\[3pt]
{\it  Department of Physics, University of Bergen,}\\
   {\it All\'egt.~55, N-5007 Bergen, Norway}
\vspace{0.6cm}

{\bf Ya.M. Shnir{\footnote{On leave of absence from the
Institute of Physics, Academy of
Sciences of Belarus, Minsk, Belarus.}}}\\[3pt]
{\it Department of Mathematics,}\\
{\it Technical University of Berlin, Germany}\\[2pt]
and \\[2pt]
{\it  Department of Physics, University of Bergen,}\\
   {\it All\'egt.~55, N-5007 Bergen, Norway}

\end{center}
\begin{abstract}
Dyon loops give a contribution to the matrix element for
light-by-light scattering that violates parity and time-reversal symmetry.
This effect induces an electric dipole moment for the electron,
of order $M^{-2}$, where $M$ is the dyon mass.
The current limit on the electric dipole moment of the electron
yields the lower mass bound $M>{\cal O}(1)~\mbox{TeV}$.
\end{abstract}

Very precise measurements achieved during the last decade
have opened up for a new approach in elementary particle physics.
According to this, evidence of new particles
can be extracted from indirect measurements of their virtual contribution
to processes at energies which are too low for direct production.
For example, the top quark mass as predicted from precision electroweak
data \cite{Ellis} agrees to within 10$\%$ with
direct experimental measurements \cite{top}.

This approach has recently been applied \cite{Rujula} for the estimation
of possible virtual monopole contributions to observables
at energies below the monopole mass.
Taking into account the violation of parity (and time-reversal symmetry)
in a theory with monopoles \cite{Tom}, the emergence of an electric
dipole moment was first pointed out by Purcell
and Ramsey \cite{PurRam}.
More recently, the effect due to monopole loop contributions
has been discussed \cite{FlamMur}.

From our point of view the estimation of the paper \cite{FlamMur}
needs reexamination. In the present paper we consider first the $P$
and $T$ violating contribution of the dyon loop to
the matrix element for light-by-light scattering.
Then the contribution of this subdiagram to the electric dipole moment
of the electron is estimated.
The experimental bound on the electron dipole moment
leads to a non-trivial bound on the dyon mass.

As was shown in \cite{KovOslShTol},
the effective Lagrangian density of QED coupled to dyons
contains in the one-loop approximation the fourth-order terms
\begin{equation}                            \label{DeltaL-dyon-e}
\Delta L  \approx
\frac{e^4}{360 \pi ^2 m^4} \bigl[
(\vecH^2 - \vecE^2)^2 +  7 (\vecH \vecE)^2 \bigr]
+ \frac{Qg (Q^2 - g^2)}{60 \pi ^2 M^4}
(\vecH \vecE) (\vecH^2 - \vecE^2),
\end{equation}
where the first term is the familiar Euler-Heisenberg term, with
$e$ and $m$ the charge and mass of the electron, whereas the second
term is a $P$ and $T$ non\-invariant dyon loop contribution,
with $M$ the dyon mass and
$Q$ and $g$ its electric and magnetic charges,
respectively{\footnote{The Dirac charge quantization condition
connects the electric charge of the electron and the magnetic charge of
a dyon: $eg \sim 1$ whereas the dyon electric charge $Q$ is not quantized.}}.
This expression yields the matrix element for low-energy
photon-photon scattering. In order to determine it,
we substitute into (\ref{DeltaL-dyon-e}) the expansion
\begin{equation}
F_{\mu \nu} (x) = \frac{i}{(2\pi)^4} \int \d^4 q
\left(q_{\mu} A_{\nu} - q_{\nu} A_{\mu} \right) e^{iqx} .
\end{equation}
Corresponding to the second term of (\ref{DeltaL-dyon-e}) we find
\begin{align}                            \label{L-repr}
&\frac{Qg (Q^2 - g^2)}{480 \pi ^2 M^4}
\int \d^4x~ \varepsilon_{\mu \nu \rho \sigma} F^{\mu \nu} F^{\rho \sigma}
F_{\alpha \beta} F^{\alpha \beta}
\nonumber\\
&=\frac{1}{(2 \pi)^{12}}
\int \d^4 q_1 \d^4 q_2 \d^4 q_3 \d^4 q_4 \delta (q_1 + q_2 + q_3 + q_4)
A_{\mu}(q_1) A_{\nu}(q_2) A_{\rho}(q_3) A_{\sigma}(q_4) \,
\widetilde M^{\mu \nu \rho \sigma},
\end{align}
where
\begin{equation}
                                   \label{M_tilde}
\widetilde M^{\mu \nu \rho \sigma}
=\widetilde M^{\mu \nu \rho \sigma}(q_1, q_2, q_3, q_4)
=\frac{Qg(Q^2 - g^2)}{60 \pi^2 M^4}
\varepsilon_{\alpha \beta}^{\phantom{\alpha \beta} \mu \nu}
q_1^\alpha q_2^\beta
\left[q_4^\rho q_3^\sigma -g^{\rho\sigma}(q_3 q_4) \right] \, .
\end{equation}
Symmetrizing this pseudotensor one obtains the $P$ and $T$ violating part
of the matrix element for light-by-light scattering.
With all momenta flowing inwards, $k_1 + k_2 + k_3 + k_4=0$,
the matrix element takes the form
\begin{eqnarray}                                   \label{Matrix}
M_{\mu \nu \rho \sigma}'
&=&{\textstyle\frac{1}{6}} \bigl[
 \widetilde M_{\mu \nu \rho \sigma}(k_1, k_2, k_3, k_4)
+\widetilde M_{\mu \rho \nu \sigma}(k_1, k_3, k_2, k_4)
+\widetilde M_{\mu \sigma \nu \rho}(k_1, k_4, k_2, k_3)
\nonumber\\
& & %\hspace{3mm}
+\widetilde M_{\nu \rho \mu \sigma}(k_2, k_3, k_1, k_4)
+\widetilde M_{\nu \sigma \mu \rho}(k_2, k_4, k_1, k_3)
+\widetilde M_{\rho \sigma \mu \nu}(k_3, k_4, k_1, k_2)
\bigr]
\nonumber\\
&=& \frac{Qg(Q^2 - g^2)}{60 \pi^2 M^4} \Bigl(
\varepsilon_{\alpha \beta }^{\phantom{\alpha \beta} \mu \nu}
k_1^\alpha k_2^\beta k_3^\sigma k_4^\rho
+ \varepsilon_{\alpha \beta }^{\phantom{\alpha \beta} \mu \rho}
k_1^\alpha k_2^\sigma k_3^\beta k_4^\nu
+ \varepsilon_{\alpha \beta }^{\phantom{\alpha \beta} \mu \sigma}
k_1^\alpha k_2^\rho k_3^\nu k_4^\beta
\nonumber\\
&&+ \varepsilon_{\alpha \beta }^{\phantom{\alpha \beta} \nu \rho}
k_1^\sigma k_2^\alpha k_3^\beta k_4^\mu
+ \varepsilon_{\alpha \beta }^{\phantom{\alpha \beta} \nu \sigma}
k_1^\rho k_2^\alpha k_3^\mu k_4^\beta
+ \varepsilon_{\alpha \beta }^{\phantom{\alpha \beta} \rho \sigma}
k_1^\nu k_2^\mu k_3^\alpha k_4^\beta \\
&&- \varepsilon_{\alpha \beta }^{\phantom{\alpha \beta} \mu \nu}
g^{\rho \sigma} (k_3 k_4) k_1^\alpha k_2^\beta
- \varepsilon_{\alpha \beta }^{\phantom{\alpha \beta} \mu \rho}
g^{\nu \sigma} (k_2 k_4) k_1^\alpha k_3^\beta
- \varepsilon_{\alpha \beta }^{\phantom{\alpha \beta} \mu \sigma}
g^{\rho \nu} (k_2 k_3) k_1^\alpha k_4^\beta
\nonumber\\
&&- \varepsilon_{\alpha \beta}^{\phantom{\alpha \beta} \nu \rho}
g^{ \mu \sigma} (k_1  k_4) k_2^\alpha k_3^\beta
- \varepsilon_{\alpha \beta }^{\phantom{\alpha \beta} \nu \sigma}
g^{ \mu \rho} (k_1 k_3)  k_2^\alpha  k_4^\beta
- \varepsilon_{\alpha \beta}^{\phantom{\alpha \beta}\rho \sigma}
g^{\mu \nu} (k_1 k_2) k_3^\alpha k_4^\beta \Bigr).
\nonumber
\end{eqnarray}
Since the interaction contains an $\varepsilon$ tensor,
the coupling between two of the
photons is different from that involving the other two,
and the familiar pairwise equivalence of the six terms does not hold.
The matrix element satisfies gauge invariance (with respect
to any of the four photons),
\begin{equation}
                                       \label{gauge-inv}
k_1^\mu\, M'_{\mu\nu\rho\sigma}(k_1,k_2,k_3,k_4)=0, \qquad \mbox{etc.}
\end{equation}
We note that the above contribution to the matrix element is proportional
to the fourth power of the inverse dyon mass,
$M'_{\mu\nu\rho\sigma}\propto M^{-4}$.
However, this result is only valid at low energies,
where the photon momenta are small compared to $M$, being obtained
from an effective, non-renormalizable theory.

The contribution of this matrix element breaks the $P$
and $T$ invariance of ordinary electrodynamics.
Thus, among the sixth-order radiative corrections
to the  electron-photon vertex there are terms
containing this photon-photon scattering subdiagram with a dyon loop
contribution (see Fig.1), that induce an electric dipole moment
of the electron\footnote{This has been noted by
I.B. Khriplovich \cite{Khrip} --- see also a recent paper by
Flambaum and Murray \cite{FlamMur}.}.
%%%%%%%%%%%%%%%%%%%%%%%%%%%%%%%%%%%%%%%%%%%%%%%%%%%%%%%%%%%%%%%%%%%%%%%%
\begin{figure}[htb]
\begin{center}
\setlength{\unitlength}{1cm}
\begin{picture}(10,6.)
\put(-6.5,-19.0)
{\mbox{\epsfysize=30.0cm\epsffile{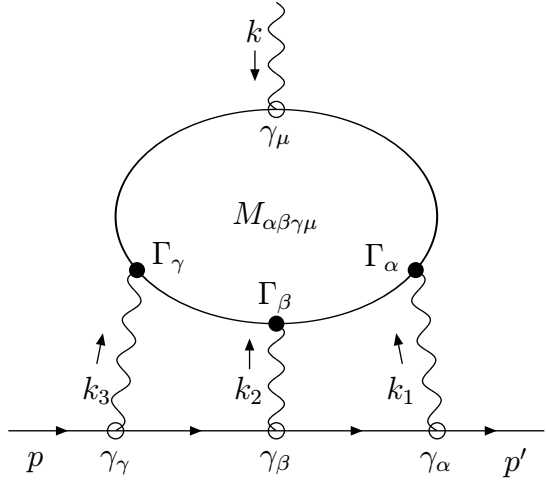}}}
\end{picture}
\caption{Typical three-loop vertex diagram.
The closed line represents a dyon loop.}
\end{center}
\end{figure}
%%%%%%%%%%%%%%%%%%%%%%%%%%%%%%%%%%%%%%%%%%%%%%%%%%%%%%%%%%%%%%%%%%%%%%%%

Indeed, one can  write the contribution of this diagram
to the electron-photon vertex as{\footnote{Of course, 
there are more diagrams.}}:
\begin{align}                         \label{vertex}
{\Lambda}_{\mu}(p',p)
&= \frac{e^2}{(2\pi)^8} \int \d^4 k_1  \d^4 k_3
\frac {1}{k_1^2 + i\epsilon }\, \frac {1}{k_2^2 + i\epsilon }\,
\frac {1}{k_3^2 + i\epsilon}~
M_{\alpha \beta \gamma \mu}(k_1, k_2, k_3, k)
\nonumber\\
&\times
\gamma^{\alpha}
\frac {\not p' + \not k_1 + m}{(p' + k_1)^2 -m^2 + i\epsilon}
\gamma^{\beta}
\frac {\not p - \not k_3 + m}{ (p - k_3)^2 - m^2 + i\epsilon }
\gamma^{\gamma}
\end{align}
where $M_{\alpha \beta \gamma \mu}(k_1, k_2, k_3, k)$
is the polarization pseudotensor representing
the dyon box diagram contribution to the photon-photon scattering amplitude,
the low-energy limit of which is given by the pseudotensor
$M'_{\alpha \beta \gamma \mu}$ of eq.~(\ref{Matrix}).

In order to extract the electric dipole moment from the general
expression (\ref{vertex}) it is convenient, according to
the approach by {\cite{Kinoshita}} to exploit the identity
\begin{equation}           \label{gauge}
M'_{\alpha \beta \gamma \mu}(k_1, k_2, k_3, k)
= -k^\nu \frac{\partial}{\partial k^\mu}
M_{\alpha \beta \gamma \nu}(k_1, k_2, k_3, k),
\end{equation}
which can be obtained upon differentiating the gauge invariance condition
of the polarization tensor [cf.\ eq. (\ref{gauge-inv})] with respect to
$k^{\mu}$.

Substituting (\ref{gauge}) into (\ref{vertex}) we can write
the $ee\gamma$ matrix element as
\begin{eqnarray}          \label{vertex-1}
M_{ee\gamma}(p',p,k)
&=& e^\mu(k) \bar u(p') \Lambda_\mu(p',p) u(p) \nonumber \\
&=& e^\mu(k) k^{\nu} {\bar u} (p') \Lambda_{\mu \nu}(p',p) u(p)
\end{eqnarray}
where $e^{\mu}(k)$ is the photon polarization vector and
\begin{align}          \label{vertex-tens}
{\Lambda}_{\mu \nu}(p',p)
= -&\frac{e^2}{(2\pi)^8}
\int \d^4 k_1  \d^4 k_3 ~\frac {1}{k_1^2 + i\epsilon}
~\frac {1}{k_2^2 + i\epsilon} ~\frac {1}{k_3^2 + i\epsilon}
~\frac{\partial}{\partial k^{\mu}}
M_{\alpha \beta \gamma \nu}(k_1, k_2, k_3, k)
\nonumber\\
&\times
\gamma^{\alpha}
\frac {\not p' + \not k_1 + m}{( p' +  k_1)^2 -m^2 + i\epsilon}
\gamma^{\beta}
\frac {\not p - \not k_3 + m}{(p -  k_3)^2 - m^2 + i\epsilon}
\gamma ^{\gamma}.
\end{align}
Since the matrix element (\ref{vertex-1}) is already proportional
to the external photon momentum $k$, one can put $k=0$
in $\Lambda_{\mu \nu}$
after differentiation to obtain the static electric dipole moment.

Then, following \cite{Kinoshita}, we note that due to Lorentz covariance
of $\Lambda_{\mu \nu}$, it can be written in the form
\begin{equation}          \label{rel-repr}
{\Lambda}_{\mu \nu}(p',p)
= \left(\tilde A g_{\mu \nu}  + \tilde B \sigma_{\mu \nu} 
+ \tilde C P_{\mu}\gamma_{\nu} + \tilde D P_{\nu}\gamma_{\mu}  
+ \tilde E P_{\mu}P_{\nu}\right)\gamma_5
+\ldots
\end{equation}
where we have omitted terms that do not violate parity,
as well as those proportional to $k_{\mu}$, and where
$\sigma_{\mu \nu}
= (\gamma_{\mu} \gamma_{\nu} - \gamma_{\nu} \gamma_{\mu})/2$,
and
$P_{\mu} = p_{\mu} + p_{\mu}'$.

Substituting this expression into the matrix element
$M_{ee\gamma}(p',p,k)$ of eq.~(\ref{vertex-1}) one can see that
there are two contributions to the $P$ violating part,
arising from the $\tilde B$ and $\tilde C$ terms.
In order to project out the dipole moment from (\ref{vertex-1}),
one has to compare eq.~(\ref{rel-repr}) with the phenomenological
expression for the electric dipole moment $d_e$ \cite{Khrip-book}:
\begin{equation}         \label{mom-phen}
M_{ee\gamma}(p',p,k)=e^\mu(k) k^{\nu}
{\bar u} (p') \frac{d_e}{2m} \, \gamma _5 \, \sigma_{\mu \nu} \, u(p),
\end{equation}
In the non-relativistic limit it corresponds to the Hamiltonian
of interaction $ -(d_e/2m) ~{\vecs}{\vecE} $.
Thus, multiplying (\ref{rel-repr}) by $\sigma_{\mu\nu} \gamma _5$
and taking the trace we have:
\begin{equation}       \label{dip-mom}
d_e = -\frac{m}{24} ~{\rm Tr}~
\left[\sigma_{\mu \nu}\gamma _5 \Lambda^{\mu \nu} \right].
\end{equation}

In order to provide an estimate of the induced electric dipole moment
we need to estimate $\Lambda^{\mu\nu}$.
The first task is to evaluate the polarization pseudotensor
$M_{\alpha \beta \gamma \mu}$ corresponding to the virtual dyon
one-loop subdiagram.
If we were to substitute for $M_{\alpha \beta \gamma \mu}$
the low-energy form $M'_{\alpha \beta \gamma \mu}$ of
eq.~(\ref{Matrix}) into eq.~(\ref{vertex}), we would obtain
a quadratically divergent integral.

On the other hand, straightforward application
of the Feynman rules in QED with magnetic charge
(see, e.g., \cite{BlagSen}) would give for the
photon-by-photon scattering subdiagram in Fig.~1\footnote{It
should be noted that the expression (\ref{Matrix}) contains
contributions from such loop diagrams with all possible combinations
of either three or one magnetic-coupling vertex $\Gamma_{\rho}$.},
\begin{align}                   \label{int-q}
M_{\alpha \beta \gamma \mu}(k_1, k_2, k_3, k)
&= \frac{Q g^3}{2\pi^4} \int \d^4 q ~{\rm Tr}~\biggl(
\Gamma_{\alpha} \frac{1}{\not q + \not k_1 - M}
\Gamma_{\beta} \frac{1}{\not q - \not k_3 - \not k - M}
\nonumber\\
& \times
\Gamma_{\gamma} \frac{1}{\not q - \not k - M}
\gamma_{\mu} \frac{1}{\not q - M} \biggr).
\end{align}
Here $\Gamma_\alpha$ represents the magnetic coupling of the photon
to the dyon, which we take according to ref.~\cite{BlagSen}
to be
\begin{equation}               \label{g-vert}
\Gamma_{\mu} = - i\varepsilon_{\mu \nu \rho \sigma}
\frac{ \gamma ^{\nu} k^{\rho} n^{\sigma}}{(n \cdot k)}.
\end{equation}
The vertex function depends on $k^{\rho}$, the photon momentum 
entering the vertex, and on $n^{\sigma}$, a unit constant space-like 
vector corresponding to the Dirac singularity line. 
It was shown by Zwanziger \cite{Zwanz}
that although the matrix element depends on $n$,
the cross section as well as other physical quantities are $n$ independent.

Calculations using this technique are very complicated and can only
be done in a few simple situations \cite{BlagSen}, for example,
in the case of the  charge-monopole scattering problem \cite{Rabl}.
We will here avoid this approach.

While the integration over $q$ in eq.~(\ref{int-q}) is logarithmically
divergent (the magnetic couplings in (\ref{int-q}) are dimensionless),
after renormalization the sum of such contributions must in the
low-energy limit reduce to the form given in eq.~(\ref{Matrix}).
We also note that the substitution of (\ref{int-q})
into eq.~(\ref{vertex-tens}) yields a convergent integral.
Thus, the following method for evaluating $\Lambda_{\mu\nu}$
suggests itself. We divide the region of integration into
two domains: (i) the momenta $k_1$ and $k_3$ are small compared
to $M$, and (ii) the momenta are of order $M$ (or larger).

In the first region, the form (\ref{Matrix}) can be used,
but since the integral is quadratically divergent, the integral
will be proportional to $M^2$.
Together with the over-all factor $M^{-4}$ this will give
a contribution $\propto M^{-2}$.
For large values of the photon momenta, the other form,
eq.~(\ref{int-q}) can be used. This gives a convergent
integral, and dimensional arguments determine the scale
to be $M^{-2}$.
It means that
\begin{equation}                  \label{estim}
\left|\Lambda_{\mu \nu}\right|
\sim \frac{ e^2 Qg(Q^2 - g^2)}{(4\pi^2)^3 M^2} .
\end{equation}
The numerical coefficient has been estimated as $1/(4\pi^2)^3$,
one factor $1/4\pi^2$ from each loop, and the
$1/24$ of eq.~(\ref{dip-mom}) is assumed cancelled by
a combinatorial factor from the number of diagrams involved.
This is of course a very rough estimate.

Now we can estimate the order of magnitude of the electron 
dipole moment generated by virtual dyons.
It is obvious from Eqs.~(\ref{dip-mom}) and (\ref{estim}),
that in order of magnitude one can write
\begin{equation}            \label{dip-mom-est}
d_e \sim \frac{e^2 Qg(Q^2 - g^2)}{(4\pi^2)^3 }\frac{m}{M^2} .
\end{equation}

This estimate can be used to obtain a new bound on the dyon mass.
Indeed, recent experimental progress in the search for an
electron electric dipole moment \cite{BernSuzuki} gives a rather strict
upper limit: $d_e < 9\cdot 10^{-28} e $ cm.
If we suppose that $Q \sim e$, from (\ref{dip-mom-est}) one can obtain
$M \geq 2\cdot 10^6~ m \approx 10^{3}$ GeV. This estimate shows
that the dyon mass belongs at least to the electroweak scale.

The above estimate coincides with the bound obtained by De R\'ujula
\cite{Rujula} for monopoles, from an analysis of LEP data, but it
is weaker than the result given in \cite{FlamMur},
where the limit $M \geq 10^{5}$ GeV was obtained. 
The authors of ref.~\cite{FlamMur} used the hypothesis that a radial
magnetic field could be induced due to virtual dyon pairs.
In order to estimate the effect, they used the well known
formula for the Ueling correction to the electrostatic potential,
simply replacing the electron charge and mass with those of the monopole.
But the Ueling term is just a correction to the scalar Coulomb potential
due to vacuum polarization and cannot itself be considered as a source
of radial magnetic field.
Indeed, there is only one second order term in the effective Lagrangian
that can violate the $P$ and $T$ invariance of the theory, namely
$\Delta L' \propto {\vecE}{\vecH}$.
But in the framework of QED there is no reason to consider such a correction
because it is just a total derivative. 
The reference to the $\theta$-term,
used in \cite{FlamMur} to estimate the electric charge of the dyon,
is only relevant in the context of a non-trivial topology
(e.g., in the 't Hooft-Polyakov monopole model) where their limit
applies. In this case there are arguments in favour of stronger
limits on the monopole (dyon) mass (see, e.g., \cite{Godd}).

Finally, one should note that the dyon loop diagram considered above
can also contribute to the neutron electric dipole moment.
The experimental value $d_n < 1.1 \cdot  10^{-25} e $ cm \cite{Altarev}
will in the naive quark model with $m \approx 10$~MeV allow us 
to obtain a similar estimate of the dyon lower mass bound
as obtained for the electron.

\medskip

One of us (Ya. S.) is very indebted, for fruitful discussions, 
to Prof.\ I. B. Khriplovich, to whom belongs the idea of 
the above described mechanism of an electric dipole
moment generated in QED with a magnetic charge.
Ya. S. is also thankful for hospitality at the University of Bergen.
This research  has been supported by the Alexander von Humboldt Foundation
in the framework of the Europa Fellowship,
and (P. O.) by the Research Council of Norway.
Ya. S. acknowledges in the first stage of this work support by the
Fundamental Research Foundation of Belarus, grant No F-094.

\end{document}